# Topological Lifshitz Transitions and Fermi Arc Manipulation in Weyl Semimetal NbAs

H. F. Yang[1,2,3*], L. X. Yang[4*], Z. K. Liu[1,2*], Y. Sun[5*], C. Chen[3], H. Peng[3], M. Schmidt[5], D. Prabhakaran[3], B. A. Bernevig[6], C. Felser[5], B. H. Yan[7] and Y. L. Chen[1,2,3,4†]

[1]*School of Physical Science and Technology, ShanghaiTech University, Shanghai 200031, P. R. China*
[2]*ShanghaiTech Laboratory for Topological Physics, 200031 Shanghai, P. R. China*
[3]*Department of Physics, University of Oxford, Oxford, OX1 3PU, UK*
[4]*State Key Laboratory of Low Dimensional Quantum Physics and Department of Physics, Tsinghua University, Beijing 100084, P. R. China*
[5]*Max Planck Institute for Chemical Physics of Solids, D-01187 Dresden, Germany*
[6]*Department of Physics, Princeton University, Princeton, New Jersey 08544, USA*
[7]*Department of Condensed Matter Physics, Weizmann Institute of Science, Rehovot 7610001, Israel*

*These authors contributed equally to this work*
† *Email: yulin.chen@physics.ox.ac.uk*

Surface Fermi arcs (SFAs), the unique open Fermi-surfaces (FSs) discovered recently in topological Weyl semimetals (TWSs), are unlike closed FSs in conventional materials and can give rise to many exotic phenomena, such as anomalous SFA-mediated quantum oscillations, chiral magnetic effects, three-dimensional quantum Hall effect, non-local voltage generation and anomalous electromagnetic wave transmission. Here, by using *in-situ* surface decoration, we demonstrate successful manipulation of the shape, size and even the connections of SFAs in a model TWS, NbAs, and observe their evolution that leads to an unusual topological Lifshitz transition not caused by the change of the carrier concentration. The phase transition teleports the SFAs between different parts of the surface Brillouin zone. Despite the dramatic surface evolution, the existence of SFAs is robust and each SFA remains tied to a pair of Weyl points of opposite chirality, as dictated by the bulk topology.

**Introduction**

Fermi-surfaces are boundaries of electronic bands in reciprocal space that separate occupied and unoccupied states. The shape and volume of FSs determine the electric, magnetic, optical and even thermal properties of materials. In conventional materials, FSs form closed shapes, such as closed curves in two-dimensional (2D) electronic systems or closed surfaces in three-dimensional (3D) electronic systems. However, this orthodoxy was recently upended by the theoretical and experimental discovery of a new type of quantum state – the topological Weyl semimetal (TWS), in which open FSs — surface Fermi arcs (SFAs) — emerge[1-5].

In a TWS, linearly dispersive bulk conduction and valence bands reminiscent of the dispersion of massless particles touch at an even number of Weyl points[1-5] (see Fig. 1a, where Weyl points in momentum space resemble magnetic monopoles). In a TWS, each Weyl point is associated with a positive or negative chirality. On the surface of a TWS, exotic topological surface states whose Fermi surfaces consist of open curves appear. The open curves are the SFAs that must originate from and terminate at the surface projection of a pair of Weyl points of opposite chirality[1-3] (see Fig. 1a).

The unusual SFAs and their interplay with bulk Weyl fermions can result in many exotic physical phenomena, such as anomalous SFA-mediated quantum oscillations[6-9] (Fig. 1b), the realization of chiral magnetic effects (e.g., the chiral anomaly[10-13] and the universal quantized contribution of SFAs in magneto-electric transport[14]), 3D quantum Hall effect[15], novel quasiparticle interference in tunnelling spectroscopy[16-19], intriguing non-local voltage generation (Fig. 1c) and unusual electromagnetic wave transmission[20-23] (Fig. 1d), and the emergence of Majorana modes when coupled with superconductivity[24]. Therefore, the capability of controlling and modifying the SFAs, especially the switching of SFAs between Weyl-points pairs (see Fig. 1a, b), are naturally desired in order to understand and manipulate these exciting and unusual phenomena.

In this work, by investigating the SFAs of NbAs, a model system from the transition-metal monophosphide (TMM) TWS family recently predicted and discovered[4,5,25-34], we were able to directly manipulate the SFAs by surface decoration. Using *in-situ* monitoring with angle-resolved photoemission spectroscopy (ARPES), for the first time, we clearly observe the evolution of SFAs through an unusual topological transition where the SFAs switch the pairs of Weyl points they connect. Remarkably, despite the dramatic change of the shape and size of SFAs during the topological transition, each SFA remains connected to a pair of Weyl points (albeit different pairs before and after the transition) of opposite chirality, demonstrating the topological nature of the electronic structures of TWSs. Notably, this topological Lifshitz transition between open Fermi surfaces is not caused by the simple shift of the Fermi-energy due to the change of the carrier concentration; rather, it is caused by a delicate surface modification with ordered decoration atoms. Furthermore, the Fermi surface consists only of SFAs unaccompanied by the kinds of topologically trivial states that are commonly observed in this family compounds[25-33]. The method to control and manipulate the topological electronic states of TWSs in this work will not only open a new frontier for the exploration of fundamental physical phenomena but also make possible the design of unusual electronic applications mediated by SFAs.

**Results**

**Basic characterization of NbAs crystals.** NbAs crystallizes into a body-centred tetragonal structure without inversion symmetry[35], with four Nb-As layers in a unit cell stacking along the c direction (Fig. 1e). The crystal can be naturally cleaved between two adjacent Nb-As layers to yield a flat (001) surface ideal for ARPES measurements. As proposed by calculations[4,5,34], there are 12 pairs of Weyl points in the Brillouin zone (BZ) connected by the SFAs (Fig. 1f), which is similar to other TWSs in the same family[25-29,31]. The high-quality NbAs crystals (Fig. 1g) synthesized for this work show sharp X-ray diffraction patterns along all three crystalline orientations, and large, mirror-like flat surfaces that facilitate the ARPES measurements. Similar to other recent reports[27,29], in order to focus on the surface states, we utilize low-energy photons (e.g. 56 eV) to maximize the surface sensitivity and increase the energy and momentum resolutions[36].

**Band structure of pristine surface.** The electronic structure of pristine, freshly cleaved NbAs (Fig. 2a, c) show the characteristic cross-shaped FS that consists of spoon- and bowtie-like surface FS pockets similar to other TWS compounds of the same family[25-29,31-33]. These FSs, whose surface nature is confirmed by photon-energy dependent ARPES measurements (see Fig. 2f and Supplementary Figure 2 and 3 for details) and extensively discussed in recent works[30], agree well with our *ab-initio* calculations of surface states on the As-terminated (001) surface (Fig. 2b). These calculations confirm the connection pattern of the SFAs as sketched in Fig. 2d.

**Figure-8-like SFAs emerging on potassium decorated surface.** To modify the SFAs and examine their topological character, we decorated the sample surface by *in-situ* potassium (K) deposition[37-39]. On one hand, surface decoration is an effective method to disrupt the normal (trivial) surface states by changing the surface environments and introducing new surface Fermi surfaces. On the other hand, as the SFAs are protected by the topology of the bulk band structures, which cannot be altered by surface modification, they have to remain even after surface decoration. Remarkably, after the surface decoration, a completely different FS emerged (Fig. 2g, i). The surface origin of the changing FS can be confirmed by photon-energy dependent ARPES measurements (see Fig. 2l, and Supplementary Figure 4 and 5). Compared to the pristine FS in Fig. 2a and 2c, the new FS differs not only in the size, but also the shape and the connection pattern of each FS pieces.

In order to understand this change, and to make connection with the experimental data, we carried out *ab initio* calculations to simulate the surface decoration process (Supplementary Figure 9 and Supplementary Note 2). The results (Fig. 2h, from a K-covered NbAs surface) reproduce the figure-8-like surface FS that agrees excellently with the experimental results (Fig. 2i). Interestingly, the calculations also show that the FSs now consist of two SFAs with a topologically different connection pattern to the Weyl points: unlike the pristine surface where the spoon-like SFAs connect the adjacent WPs (Fig. 2d) within the same BZ, the SFAs after decoration now even connect WPs from different BZs (see Fig. 2j). This is a manifestation of a topological quantum phase transition. Remarkably, despite the dramatic change of their shape and size, the SFAs still abide by the rule that each SFA must initiate and terminate at Weyl points of opposite chirality, as shown by stacking-plots of measured surface FSs and bulk dispersions with their momentum aligned in Fig. 2e and 2k. This clearly demonstrates the topological robustness of the SFAs (bulk states measurements and detailed discussion can be found in Supplementary Note 4).

Furthermore, unlike the pristine surface where the FS consists of both SFAs (marked as SFAs) and conventional surface states (e.g. the bowtie-like surface states, marked as Trivial SS in Fig. 2a, 2c and previous works[25-29, 31-33]), the FS after the surface decoration also shows a much simpler structure, which only consists of two long SFAs without other surface states (Fig. 2g, i) and makes it the cleanest system of Fermi-arcs observed to date. This simplicity, as well as the rather large length of SFAs, will help realize SFA-related phenomena (e.g. novel quantum oscillations, non-local transport, quasiparticle interference in tunneling spectroscopy, Majorana modes, etc.) without the spurious effects of trivial surface states.

In addition to the FSs, the difference between surface electronic structures before and after the surface modification can be seen in other aspects of the electronic structure, such as the band dispersions and the evolution of the constant energy contours, as demonstrated in Fig. 3. The accumulation of the dosed K atoms on the sample surface can be clearly seen by the characteristic K $3p$ level, which is absent on the pristine surface (Fig. 3a) but emerges after the surface decoration (Fig. 3e). Associated with the surface modification, we summarize the overall band structure in volume plots (Fig. 3b, f), the high-symmetry band dispersions across the whole BZ (Fig. 3c, g), and the evolution of band contours at different binding energies (Fig. 3d, h). Notably, the side-by-side comparisons of our comprehensive electronic structures studies show excellent agreement with the surface states' band structure results from our *ab initio* calculations.

Besides the agreement of the general band structure, in Fig. 4 the fine structure of the figure-8-like FS is further resolved. 2D curvature analysis[40] results (Fig. 4b), Fermi contours at higher binding energies (like $E_B$ = 0.1 eV in Fig. 4d), and band dispersions for cuts through the figure-8 showing four band crossings of the topological surface states (TSSs) (Fig. 4e), consistently reveal that the figure-8-like FS is composed of two long SFAs, and confirms their connectivity to projections of Weyl points as predicted (Fig. 4c).

**Topological quantum Lifshitz transition.** To investigate the evolution of the SFAs and its topological phase transition, we carried out detailed K-dose dependent measurements of the band structure, as illustrated in Fig. 5. The overall evolution constitutes of two stages: Stage I is from Panel a – c and Stage II is from d – f of Fig. 5. Within each stage, the surface band structure including the FS and dispersion evolve continuously and show only quantitative changes (such as the continuously shrinking of the spoon- and bowtie-like FS pockets in Stage I and the growing of the figure-8-like FS pieces in Stage II). However, between the two stages, a topological quantum Lifshitz transition (TQLT) occurs, manifested by the dramatic change of the FS topology and the band dispersions, which is concomitantly accompanied by the disappearance of the bowtie-like trivial SSs in stage I, leaving only the SFAs of Stage II. Such a transition cannot be explained by a simple rigid band shift of $E_F$, as is also confirmed by our *ab initio* calculations (more details can be found in Supplementary Note 2). Further orbital calculations show that while the TSS of the pristine surface harbours a complicated multi-orbital contributions from both As and Nb, the TSS of K-decorated is dominated by Nb $d_{z^2}$ orbital (Supplementary Figure 8).

**Discussion**

We have proved the capability of direct manipulation of the SFAs and the control of their topological character: the topologically distinct connection patterns between the Weyl points. For the first time, we have observed a Lifshitz transition of an open Fermi surface; provided an important, yet previously unobserved proof that Fermi arcs are topological; and shown how to change the topology of a surface Fermi arc. By varying the surface physical properties (e.g. the carrier concentration), we provided an effective method to investigate the unique topological electronic structures of TWSs, with future work needed to explore more of their exotic physical phenomena and, possibly, unusual electronic applications.

**Methods**

**Synthesis of bulk single crystals.** NbAs single crystals were synthesized by chemical vapor transport using iodine as the transport agent. We first obtained a polycrystalline precursor, then grew single crystals from it. Precursor polycrystalline samples were prepared from a mixture of high-purity (>99.99%) Nb and As elements. The mixture was sealed under high vacuum in a quartz tube, which was sealed in another evacuated tube for extra protection. First, the vessel was heated to 600°C at a rate of 50°C/hour. After 10 hours' soaking, it was slowly heated to 1050°C at a rate of 30°C/hour and then kept at this temperature for 24 hours. Finally the vessel was cooled to room temperature. From the resulting polycrystalline precursor, single crystals were grown using chemical vapor transport in a two-zone furnace. The polycrystalline NbAs powder and 0.46 mg/cm$^3$ of iodine were loaded into a 24-mm-diameter quartz tube and sealed under vacuum. The charged part of the tube was kept at 1150°C and the other end was kept at 1000°C for three weeks. The resulting crystals with shiny mirror-like surfaces can be as large as 0.5 - 1.0 mm in size, suitable for standard high-resolution ARPES measurements.

**Angle-resolved photoemission spectroscopy (ARPES).** ARPES measurements were performed at BL I05 of the Diamond Light Source (DLS), UK. Data were recorded by a Scienta R4000 analyzer. The measurement sample temperature and pressure were 10 K and 1.5×10$^{-10}$ Torr, respectively. The overall energy and angle resolutions were 15 meV and 0.2°, respectively. Linearly horizontally polarized (LH) photons are used as they show high intensity with all band structure features being clearly captured. Fresh NbAs surfaces for ARPES measurements were obtained by cleaving in-situ the crystals at low temperatures (10 K). Potassium atoms were evaporated in-situ onto cleaved (001) NbAs surfaces by a commercial SAES potassium dispenser under an ultrahigh vacuum of $2.0\times10^{-10}$ Torr.

*ab initio* **calculations.** Electronic structures were calculated by the density-functional theory (DFT) method which is implemented in the Vienna *ab initio* Simulation Package (VASP)[41]. The core electrons were represented by the projected augmented wave method[42]. The exchange-correlation was considered in the generalized gradient approximation (GGA)[43] and spin-orbital coupling (SOC) was included self-consistently. The energy cut off was set to be 300 eV for the plane-wave basis. Experimental lattice parameters were used in the construction of a model slab with a thickness of seven unit cells to simulate a surface, in which the top and bottom surfaces are terminated by As and Nb, respectively. Positions of outermost four atomic layers were fully optimized to consider surface atomic relaxation. The surface band structures and Fermi surfaces were projected onto the first unit cell of the As-terminated side, which fit the experimental band structure well. We adopted $12\times12$ and $400\times400$ k-point grids in the self-consistent charge and in the Fermi surface calculations, respectively.

**Data availability.** The data generated and analysed during this study are available from the corresponding author upon request.

**Acknowledgements**
We thank John A. McGuire for polishing the writing of this paper. This work was supported by the National Key R&D program of China (No. 2017YFA0305400), the National Natural Science Foundation of China (Grants No. 11774190, No. 11674229 and No. 11634009). Y.L.C. acknowledges insightful discussion with Siddharth Parameswaran and the support of the EPSRC Platform Grant (Grant No EP/M020517/1) and Hefei Science Center CAS (2015HSC-UE013). Y.L.C. and L.X.Y. acknowledge the support from Tsinghua University Initiative Scientific Research Program. B.A.B. acknowledges an early discussion with Ady Stern in Banff, 2015, where he was introduced to the idea of changing the Fermi arc topology of a Dirac semimetal. B.A.B. was supported by Department of Energy de-sc0016239, Simons Investigator Award, the Packard Foundation, and the Schmidt Fund for Innovative Research, NSF EAGER grant DMR-1643312, ONR-N00014-14-1-0330, ARO MURI W911NF- 12-1-0461, NSF-MRSEC DMR-1420541. C.F. acknowledges financial support by the ERC Advanced Grant (No 291472 'Idea Heusler'). B.Y. is supported by a research grant from the Benoziyo Endowment Fund for the Advancement of Science and by a Grant from the German-Israeli Foundation for Scientific Research and Development (GIF Grant no. I-1364-303.7/2016). H.F.Y. acknowledges financial support from the China Postdoctoral Science Foundation (Grant No. 2017M611635).

**Author contributions**
Y.L.C. conceived the experiments. H.F.Y., L.X.Y. and Z.K.L. carried out ARPES experiments with the assistance of C.C. and H.P. M.S., D.P. and C.F. synthesized and characterized bulk single crystals. Y.S. and B.H.Y. performed *ab initio* calculations. B.A.B contributed to the scientific discussion and revision of the paper.

**Additional information**
Competing interests: The authors declare no competing interests.

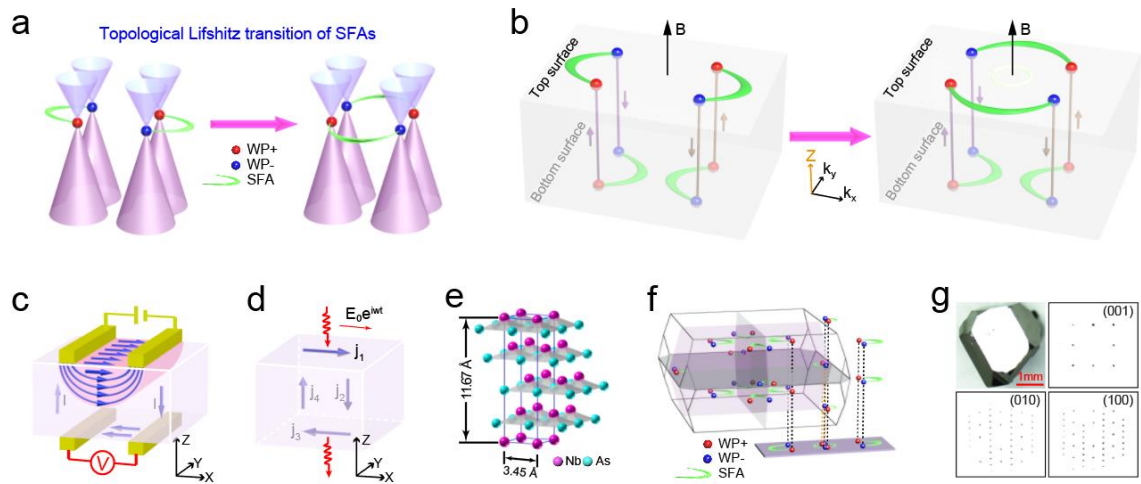

**Figure 1 | Topological Lifshitz transition of SFAs and characterization of NbAs crystals. a**, Sketch of topological Lifshitz transition of SFAs in TWSs. Balls and green crescents represent Weyl points and Fermi arcs, respectively. **b**, Schematic illustration of Weyl orbits that weave together chiral bulk states and SFAs of top and bottom surfaces in magnetic fields. When a topological Lifshitz transition of SFAs occurs at the top surface, the Weyl orbits and relevant quantum phenomena change, thus modifying the physical responses of TWSs. **c**, Schematic illustration of non-local voltage generation[21]. **d**, Schematic illustration of unusual electromagnetic wave transmission[21]. **e,** Sketch of the NbAs crystal structure, showing a body-centered-tetragonal structure without inversion symmetry. **f**, Bulk and projected surface Brillouin zones (BZs) with Weyl points connected by surface Fermi arcs. **g**, Optical image of a typical NbAs crystal and X-ray diffraction patterns along the (001), (010), and (100) crystalline directions.

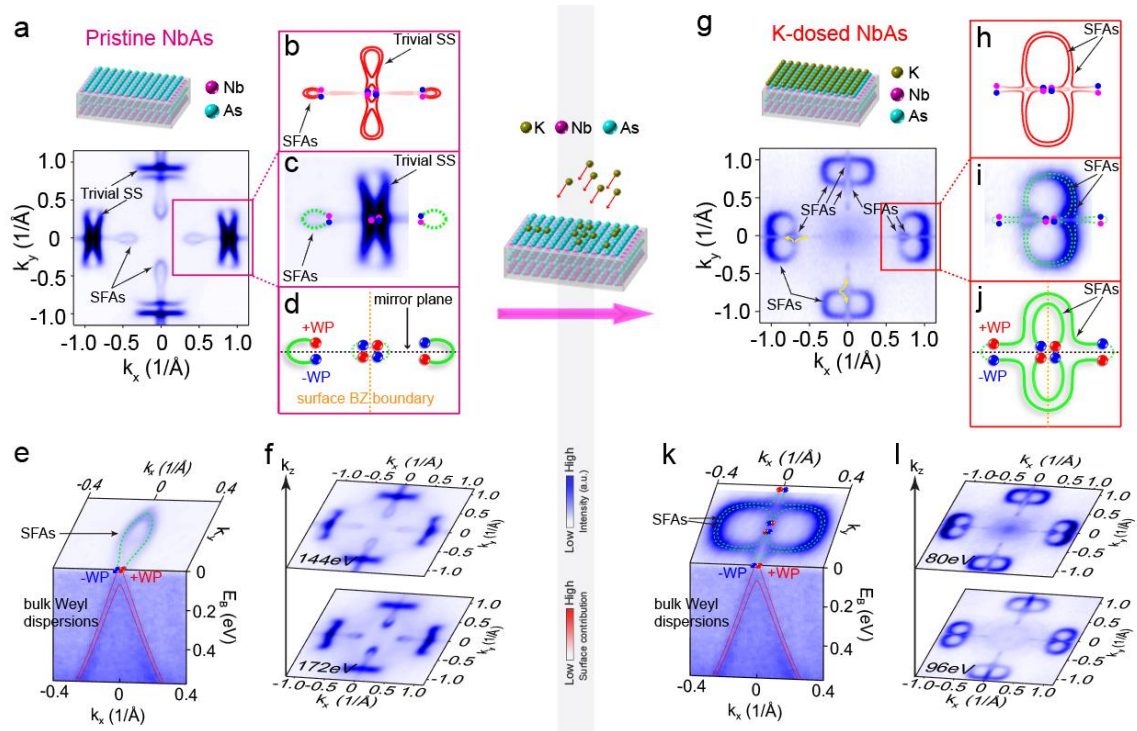

**Figure 2 | Fermi surface comparison of pristine and potassium (K)-decorated NbAs. a**, Fermi surface of pristine NbAs consisting of SFAs and trivial surface states (marked as Trivial SS). **b-c** are the calculated and measured Fermi surface around the $\bar{X}$ point, respectively. **d**, Sketch of the SFAs' connectivity to projections of Weyl points. **e**, In pristine NbAs, SFAs connect projections of bulk Weyl points (upper image: surface Fermi surface showing the SFAs; lower image: bulk Weyl dispersions with overlaid calculated bulk dispersion). **f**, two representative Fermi surfaces measured with 144 and 172 eV photons show spoon- and bowtie-like features of almost identical size, thus revealing their surface nature. **g**, Figure-8-like Fermi surface of K-decorated NbAs consisting of long SFAs without trivial surface states. The overlaid yellow arrows show that the figure-8-like FS splits into two branches near $k_x= 0$ and $k_y = 0$ regions. **h-i** are the calculated and measured Fermi surface around the point, respectively, which show good agreement. **j**, The SFAs and connection pattern are schematically displayed. **k**, In K-decorated NbAs, SFAs connect projections of bulk Weyl points (upper image: surface Fermi surface showing the SFAs; lower image: bulk Weyl dispersions with overlaid calculated bulk dispersion). **i,** Two representative Fermi surfaces measured with 80 and 96 eV photons, show figure-8-like features of almost identical size, thus revealing their surface nature. Horizontally polarized photons are used as they produce high intensity signals and clearly reveal all band structure features. (Note that the color scale of the calculated bands represent their surface

projection, i.e. the bold red curves represent the surface states while the faded red curves represent bands with strong bulk origin).

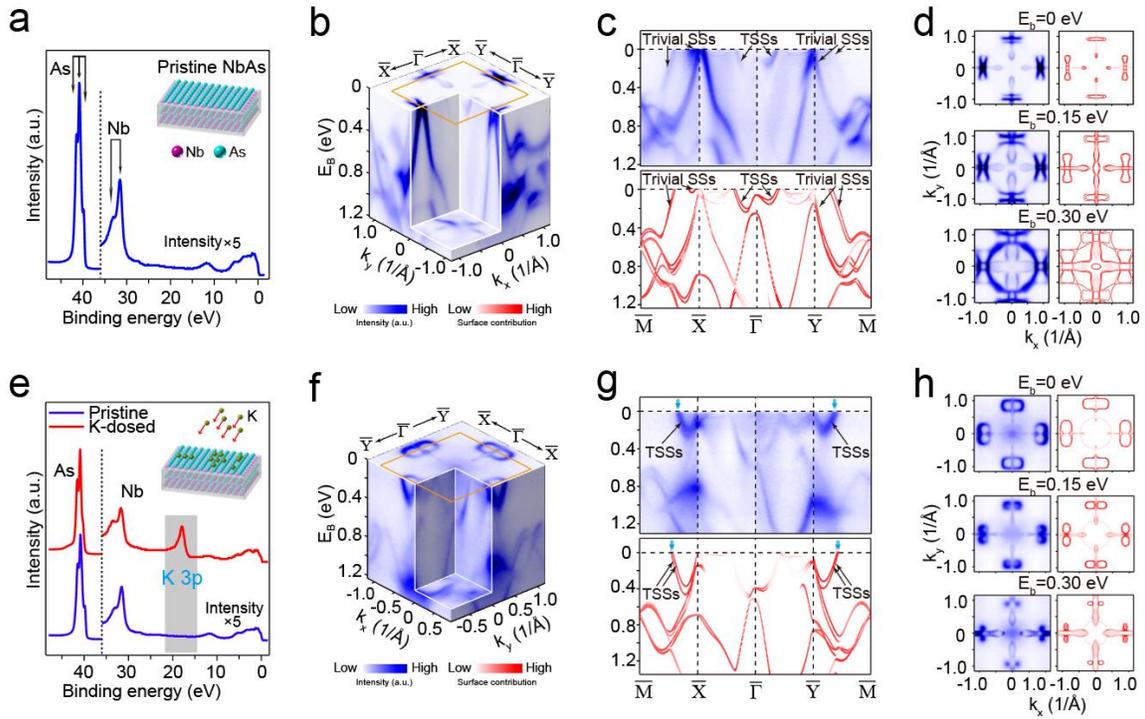

**Figure 3 | Measured and calculated band structures of pristine and K–decorated NbAs. a**, Core-level photoemission spectrum of cleaved NbAs (001) surface (pristine), clearly showing the characteristic As and Nb peaks. **b**, 3D intensity plot of the band structure illustrating the band dispersions. **c**, Excellent agreement between measured and calculated band dispersions along high-symmetry directions with trivial surface states (Trivial SSs) and topological surface states (TSSs generating SFAs) marked. **d**, Side-by-side comparison between experiments and calculations, which exhibits good agreement, of three representative constant-energy contours ($E_b$ = 0, 0.15, 0.30 eV). For comparison, these contours are obtained by mirroring the photoemission spectrum with respect to the $k_y$ axis. **e**, Core-level photoemission spectrum of K-dosed NbAs, which clearly displays the K *3p* peak, and pristine NbAs. **f** – **h**: same as **b** – **d** but displaying the band structures of K-decorated NbAs. Our measured data agree well with the calculations. In **g**, Cyan arrows mark band crossings of the TSSs (corresponding to SFAs). (Note that the color scale

of the calculated bands represent their surface projection, i.e. the bold red curves represent the surface states while the faded red curves represent bands with strong bulk origin).

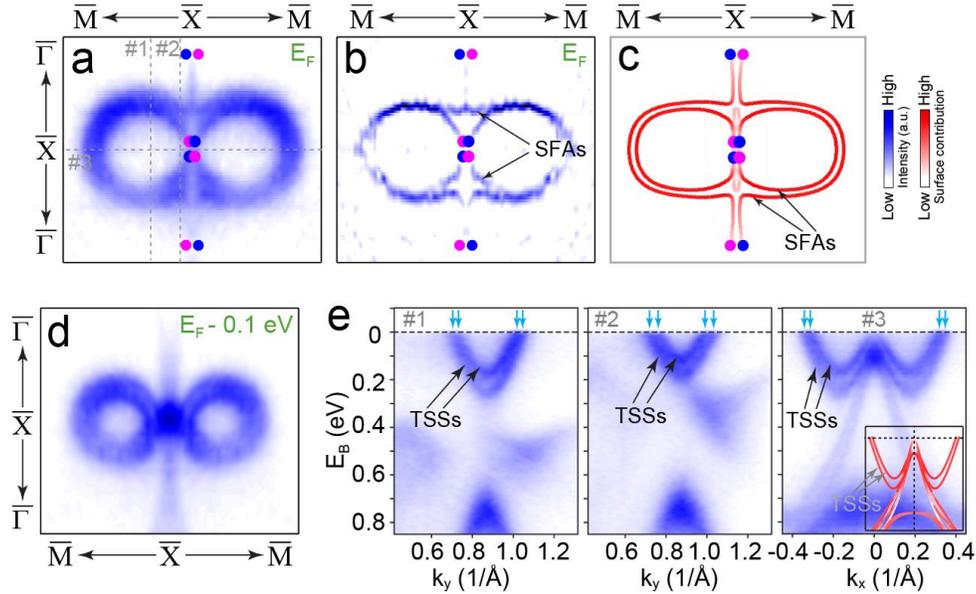

**Figure 4 | Resolving the fine structure of figure-8-like FS. a-c** are the measured Fermi surface patch around the $\overline{X}$ point, 2D curvature analysis results and calculations, respectively. Nominal K dosage for this set of data is 1.8 monolayer (ML), which is slightly smaller than that of the K-dosed data presented in Figs. 2 and 3. **d**, Measured constant-energy contour at 0.1 eV below $E_F$. **e**, Three representative band dispersions along cut #1, #2, and #3, clearly showing four band crossings of the TSSs that leads to two long SFAs forming the figure-8-like FS.

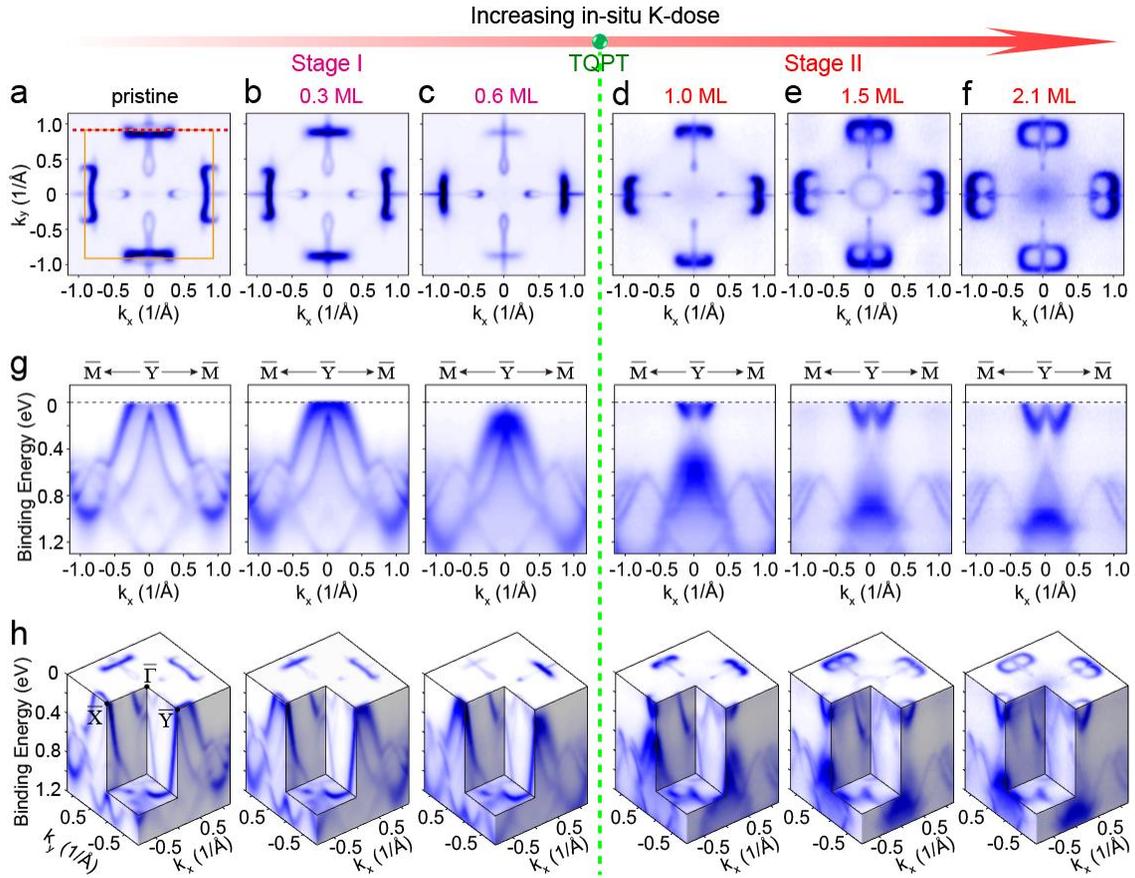

**Figure 5 | Detailed band structure evolution of topological Lifshitz transition of SFAs. a – f**, **g**, and **h** are the evolution of the FS, representative band dispersion along the $\overline{\mathrm{M}} - \overline{\mathrm{Y}} - \overline{\mathrm{M}}$ direction, and 3D intensity plot of the overall band structure with progressively increasing the K dose, respectively. For comparison, these contours are obtained by mirroring the photoemission spectrum with respect to the $k_y$ axis. The overall evolution consists of two stages: Stage I is from Panel **a** to Panel **c**, and Stage II is from Panel **d** to Panel **f**. Between these two stages, a surface topological quantum Lifshitz transition (TPLT) occurs. We note that the uneven intensity of the FS between the first and second BZs are due to the photoemission matrix element effect[36].